\documentclass[12pt,preprint]{aastex}

\newcommand{\nn}{\nonumber}

\shorttitle{Particle acceleration around shocks IV}
\shortauthors{Morlino et al.}

\begin{document}

\title{Particle acceleration at shock waves: particle spectrum as a
       function of the equation of state of the shocked plasma}

\author{G. Morlino}
\affil{Dipartimento di Fisica, Universit\`a di Pisa,
    Pisa, Italy}

\author{P. Blasi}
\affil{INAF/Osservatorio Astronomico di Arcetri, Firenze, Italy}

\and

\author{M. Vietri}
\affil{Scuola Normale Superiore, Pisa, Italy}

\begin{abstract}
We determine the spectrum of particles accelerated at shocks with
arbitrary speed and arbitrary scattering properties for different
choices of the equation of state of the downstream plasma. More
specifically we consider the effect of energy exchange between the
electron and proton thermal components downstream, and the effect of
generation of a turbulent magnetic field in the downstream plasma. 
The slope of the spectrum turns out to be appreciably affected by all
these phenomena, especially in the Newtonian and trans-relativistic
regime, while in the ultra-relativistic limit the universal spectrum
$s\approx 4.3$ seems to be a very solid prediction.
\end{abstract}


\section{Introduction}
\label{sec:intro}

The mechanism of diffusive particle acceleration at shock fronts
is one of the best studied acceleration processes possibly at
work in a wide variety of astrophysical environments. After the
pioneering work carried out in the 70's and early 80's on
non-relativistic shock fronts \cite{krymskii,bo78,bell78a,bell78b}
and slightly later on the extension of the acceleration process to the
case of relativistic shocks \cite{peacock,heavens} numerous
developments took place mainly driven by three scientific goals: 1)
relax the assumption of {\it test particles}, therefore introducing
the possibility of shock modification due to the dynamical reaction of
the accelerated particles on the shock itself (see \cite{drurymalkov}
for a recent review); 2) introduction of self-generation of scattering 
as a result of streaming instability; 3) set the bases for a general 
theory that could be applied to shocks with arbitrary speed. 

From the phenomenological point of view these efforts are motivated by
the wealth of situations in which the acceleration process can
potentially be at work, from the case of supernova remnants (SNRs) to
that of gamma ray bursts and more in general to the case of relativistic
jetted plasma ejections from compact objects. The case of SNRs is
especially relevant for its potential implications for the origin of
galactic cosmic rays. 

The point 1) above has received much attention: the problem of
dynamical reaction of the accelerated particles has been faced 
by using two-fluid models \cite{dr_v80,dr_v81}, kinetic models 
\cite{malkov1,malkov2,blasi1,blasi2,vannoni,amato1} and numerical
approaches, both Monte Carlo and other simulation procedures 
\cite{je91,bell87,elli90,ebj95,ebj96,kj97,kj05,jones02}. All these
approaches are however limited to non-relativistic shocks. The only
case that we are aware of in which shock modification has been calculated 
for a relativistic shock is the numerical simulation of
\cite{double}. Recently \cite{amato2} have presented an
semi-analytical solution of the combined problem of particle
acceleration at modified shocks with self-generated Alfven waves
(point 2), though treating the wave amplification in the context of
quasi-linear theory.  

The point 3) above, namely the construction of a unified description
of the acceleration process applicable to any shock, has been
successfully solved, in the framework of a test particle approach, by
\cite{vie03,bla05}, and applied to some cases of interest by \cite{mor06}. 
\cite{bla05} illustrated a wide variety of comparisons of the theory
with existing work in both cases of non-relativistic and
ultra-relativistic shocks. In the latter case, much work has been
done by several authors, who used numerical methods 
\cite{bed98,ach01,lem03,nie04,lem06} or semi-analytical approaches
\cite{kir87,gal99,kir00}, though limited to specific situations (for
instance small pitch angle scattering).

This paper is the third of a series, in which the original calculations
of \cite{vie03,bla05} are applied to investigate cases that can be of
some phenomenological relevance. Here we determine the changes to the
spectral index of the accelerated particles, taking into account the
possibility that different physical situations may change the equation
of state of the plasma downstream of the shock. 
In particular collective plasma effects in the proximity of a
collisionless shock may be responsible for different levels of
thermalization of the electron and proton gas components behind the
shock. These effects are expected to provide a channel of energy
transfer from protons to electrons (see for example
\cite{beg88,hos92,gal92}) though not through particle-particle
collisions. Other collective processes might result in the generation
of a downstream turbolent magnetic field, which in turn may change the
equation of state of the downstream fluid. We describe these effects
in a phenomenological way, by introducing a parametrization of
the energy exchange between electrons, protons and magnetic field
energy densities in the downstream plasma, and we use such
parametrization to infer the changes in the equation of state and the
spectral slope of the particles accelerated at the shock front.

The paper is organized as follows: in \S \ref{sec:jump} we describe
the jump conditions for a strong shock moving with arbitrary speed. In
\S \ref{sec:state} we obtain the equation of state for the downstream
plasma for different cases of interest. We present our results on the
spectrum of accelerated particles in \S \ref{sec:spectrum} and
conclude in \S \ref{sec:conclusions}.

\section{Relativistic jump conditions for strong shocks}
\label{sec:jump}

The jump conditions describing the conservation of mass, momentum and
energy across a shock front moving with velocity $V_{sh}=\beta_{sh} c$
in a medium with density $n_1$, pressure $p_1$ and energy density
$\epsilon_1$ are as follows (\textit{e.g.} \cite{kir99,gal02}):
\begin{eqnarray} 
\Gamma_1 \beta_1 n_1 \hspace{2cm} &=&  \Gamma_2 \beta_2 n_2
\label{sez2eq_jump_1a}	\\ 
\Gamma_1^2 \beta_1 (\epsilon_1 + p_1) \hspace{0.9cm} &=&  \Gamma_2^2
\beta_2 (\epsilon_2 + p_2)  \label{sez2eq_jump_2a}	\\ 
\Gamma_1^2 \beta_1^2 (\epsilon_1 + p_1)+p_1 &=& \Gamma_2^2 \beta_2^2
(\epsilon_2 + p_2)+p_2 \, . \label{sez2eq_jump_3a} 
\end{eqnarray} 
Number densities ($n$), pressures ($p$) and energy densities
($\epsilon$) are all measured in the comoving frame of the plasma they
refer to, while the Lorentz factor $\Gamma_1$ ($\Gamma_2$) of the
upstream (downstream) plasma are measured in the shock frame (clearly
$\Gamma_1=\Gamma_{sh}=(1-\beta_{sh}^2)^{-1/2}$). The indexes `1' and
`2' refer to the upstream and downstream plasmas respectively.  

The equations above can be easily generalized to the case of presence of
non negligible magnetic fields upstream, but in the following we shall
assume that 
the dynamical role of such fields is always fully negligible, and
shall therefore ignore the corrections in the conservation equations. 

The system of equations Eqs. (\ref{sez2eq_jump_1a})-(\ref{sez2eq_jump_3a}) 
can be solved once an equation of state for the plasma has been fixed,
in the very general form $\epsilon=\epsilon(n,p)$. For simplicity, in
the following we shall limit ourselves with the case of strong shocks,
namely shock waves propagating in cold pressureless media, so that
$p_1=0$ and $\epsilon_1 \simeq n_1 mc^2 \equiv \rho_1 c^2$. In this
framework Eqs. (\ref{sez2eq_jump_2a}) and (\ref{sez2eq_jump_3a})
become: 

\begin{eqnarray}
\Gamma_1^2 \beta_1 \rho_1 c^2 &=&  \Gamma_2^2 \beta_2 (\epsilon_2 + p_2)
\label{sez2eq_jump_2b} \qquad {\rm and} \\ 
\Gamma_1^2 \beta_1^2 \rho_1 c^2 &=& \Gamma_2^2 \beta_2^2 (\epsilon_2 +
p_2)+p_2 \, . \label{sez2eq_jump_3b} 
\end{eqnarray} 

We assume that the equation of state has the form $\epsilon_2 = \rho_2
c^2 F(p_2/\rho_2 c^2)$, or in terms of the normalized variables $\bar
p_2=\frac{p_2}{\rho_2 c^2}$ and $\bar\epsilon_2 = \epsilon_2/\rho_2 c^2$:

\begin{equation}
\bar\epsilon_2 = F(\bar p_2).
\end{equation}

Most cases of astrophysical interest are well described by this
functional form for the equation of state of the downstream gas, as
discussed in Sec. \ref{sec:state}.

Using the equation of mass conservation, Eq. (\ref{sez2eq_jump_1a}),
Eq. (\ref{sez2eq_jump_2b}) becomes

\begin{equation} \label{sez2eq_Gamma1}
\Gamma_2 = \Gamma_1 / (\bar\epsilon_2(\bar p_2) + \bar p_2) \equiv g_1(\bar p_2),
\end{equation} 
while  from (\ref{sez2eq_jump_3b}) we have
\begin{equation} \label{sez2eq_Gamma2}
\Gamma_2^2 = (\bar\epsilon_2^2 - 1)/(\bar\epsilon_2(\bar p_2)^2 - \bar
p_2^2 -1) \equiv g_2^2(\bar p_2) \, . 
\end{equation} 
The solution for $\bar p_2$ can be obtained solving numerically the equation
$g_1(\bar p_2)= g_2(\bar p_2)$. Once $\bar p_2$ is known, the equation of state
gives $\bar \epsilon_2$ while Eq. (\ref{sez2eq_Gamma1}), or equivalently Eq.
(\ref{sez2eq_Gamma2}), gives $\Gamma_2$. Finally Eq. (\ref{sez2eq_jump_1a})
gives the number density $n_2$. At this point it is also easy to
determine the velocity jump and the shock $r_\beta=\beta_1/\beta_2$,
which is a crucial parameter for the description of the process of
particle acceleration at the shock front. 

\section{Equations of state for the downstream plasma} \label{sec:state}

In this section we consider several instances of equations of state for
the downstream gas, in addition to the well known and widely used 
Synge equation of state \cite{syn57}. In \S\ref{sez3.1} we discuss the
case of a downstream plasma made of two independent particle species 
that may thermalize to different temperatures. In \S\ref{sez3.2} we
introduce the possibility that the the proton and electron components
are coupled in a collisionless way. In \S\ref{sez3.3} we discuss the
modification of the equation of state due to generation of a turbulent
magnetic field in the downstream plasma. Finally in \S\ref{sez3.4} we
make an attempt to consider the most general case in which all the
effects described above are taken into account. 

\subsection{The case of a plasma with independent particle species} 
\label{sez3.1}

An equation of state which is widely used in the literature was
introduced by \cite{syn57}. The basic assumption is that the plasma
consists of a single component with temperature $T$, and 
\begin{equation} \label{sez3eq_Synge_single}
\epsilon + p = \rho c^2\, G \left( mc^2 / k_{\rm B} T \right)
\end{equation} 
where $G(x)=K_3(x)/K_2(x)$ and $K_2$, $K_3$ are the modified Bessel
functions. For $x\gg 1$ (\textit{i.e.} $k_{\rm B} T \ll mc^2$) Eq. 
(\ref{sez3eq_Synge_single}) reduces to the classical Newtonian 
equation of state, $\epsilon = \rho mc^2+3p/2$, while in the opposite 
limit $x\ll1$ (\textit{i.e.} $k_{\rm B} T \gg mc^2$) the
ultra-relativistic equation of state is recovered, $\epsilon=3p$. 

If the downstream plasma can be well described as an ideal gas 
($p=n k_{\rm B} T$) made of a single component, Eq. 
(\ref{sez3eq_Synge_single}) can be rewritten in terms of the
normalized variables introduced in the previous section:
\begin{equation} \label{sez3eq_Synge_single_Norm}
\bar\epsilon = G \left( 1/ \bar p \right) - \bar p,
\end{equation} 
which has the functional form assumed in \S\ref{sec:state} with $F(\bar
p)=G \left( 1/ \bar p \right) - \bar p$. The Synge equation of state
describes correctly the behaviour of the plasma in the
ultra-relativistic and newtonian limits: applying the procedure
illustrated in \S\ref{sec:jump} we easily find the velocity jump
$r_\beta$, which is plotted as a solid line in Fig. 
\ref{sez3fig_Eq_stato_p+e}, as a function of the product $\Gamma_{sh}
\beta_{sh}$. 
In the case of strong Newtonian shocks the well known result
$r_\beta=4$ is recovered. In the limit of an ultra-relativistic shock, 
when the downstream fluid obeys the ultra-relativistic equation of
state for the gas, $r_\beta = 3$. 

The generalization of the Synge equation of state to the case of two
(or more) independent particle species with temperatures $T_i$ is rather
straightforward. Of particular interest is the case in which the
temperature of the $i-$th species is simply due to the isotropization
of the velocity vectors at the shock surface. In this case the energy
density of the $i-$th species in the downstream plasma can be written
as $\epsilon_{2(i)} = \Gamma_{\rm rel} n_i m_i c^2$, or, in terms of
dimensionless variables:  
\begin{equation} \label{sez3eq_termic_conversion}
\bar\epsilon_{2(i)} = \Gamma_{\rm rel},
\end{equation} 
independent of the type of particles. Since the normalized energy
density is the same for all species, the normalized pressures need to
be the same too. It follows immediately that for the system as a whole
one can write:
\begin{equation} \label{sez3eq_Synge_multi}
\bar\epsilon_{\rm T} + \bar p_{\rm T} = \frac{1}{\rho_{\rm T}c^2}
\sum_i \rho_i c^2 \, G \left( m_ic^2 / k_{\rm B} T_i \right) =
G(1/\bar p_{\rm T}) \, , 
\end{equation} 
where the total quantities have the subscript $_T$. 

\subsection{Coupling between thermal protons and thermal electrons}
\label{sez3.2}

The formation of collisionless shocks, both in the relativistic and
newtonian regime still represents a subject of active investigation,
in that the mechanisms that allow for an efficient transport of
information among the particles in the plasma through the exchange of
MHD waves are poorly known. On the other hand we know that such shocks
do exist, which can be interpreted as an indirect proof of the
importance of collective effects in collisionless plasmas. The same
type of effects may also be responsible for total or partial
thermalization of the different components of a plasma. Whether
electrons and protons downstream of the shock front are in thermal
equilibrium or not is a matter of debate. Most likely the answer
depends on the specific conditions behind the shock of interest. 
In addition to the thermalization of the species, there are several
other problems related to our ignorance of the complex physics that
rules these effects: for instance even the spectrum of the {\it thermal}
distribution of protons (and respectively of electrons) may not be a
typical Maxwellian, in particular if a background magnetic field
makes the distribution of energy in the waves anisotropic. In these 
conditions one should probably introduce a plasma temperature along 
the field and perpendicular to it. The problem of the thermalization
of the plasmas around collisionless shocks is also related to the
issue of the {\it thickness} of a collisionless shock, which is
usually assumed to be of the order of the gyration radius of the
thermal proton component. The same collective effects also determine
the efficiency of injection of particles in the acceleration cycle: it
appears intuitively clear that in a collisionless shock the processes
of thermalization and particle acceleration to non-thermal energies
are intimately related to each other. 

Lacking a true theory of collisionless energy transport, we adopt here
a phenomenological approach in that we parametrize the degree of
equilibration between electron and proton temperatures in the
downstream region by introducing a parameter $\xi_e$, such that the
temperatures of electrons and protons satisfy the relation $T_e =
\xi_e T_p$. On very general grounds we expect $\xi_e <1$, at least
close to the shock, before any collisional effects may possibly
equilibrate the two temperatures where the plasma has moved away from
the shock front. 

If we assume that the electron and proton gas separately behave as
perfect fluids, the pressures of the two components are related
through $p_e = n k_{\rm B}T_e = \xi_e p_p$. The equations of state of
electrons and protons are easily found to be

\begin{eqnarray}
\bar\epsilon_p &=& \bar\rho_p \, G\left(\frac{\bar\rho_p}{\bar p_p} 
\right) - \bar p_p   \label{sez3eq_Energy_p}  \\ 
\bar\epsilon_e &=& \bar\rho_e \, G\left(\frac{\bar\rho_e}{\xi_e \bar p_p}
\right) - \xi_e \bar p_p \, .
\label{sez3eq_Energy_e} 
\end{eqnarray} 

Here all the normalized quantities refer to the total matter density,
namely $\bar x=x/n(m_e+m_p)c^2$. In order to solve the equations for
the jump conditions at the shock surface we also need the total
normalized energy densities and pressures:

\begin{eqnarray}
\bar\epsilon_{\rm T}(\bar p_p) &=& \bar\epsilon_p(\bar p_p) +
\bar\epsilon_e(\bar p_p)   \label{sez3eq_Energy_p+e} \\ 
\bar p_{\rm T}(\bar p_p) &=& \bar p_p (1+\xi_e)  \,. 
\label{sez3eq_Pressure_p+e} 
\end{eqnarray} 

\begin{figure}
\begin{center}
\includegraphics[angle=0,scale=1]{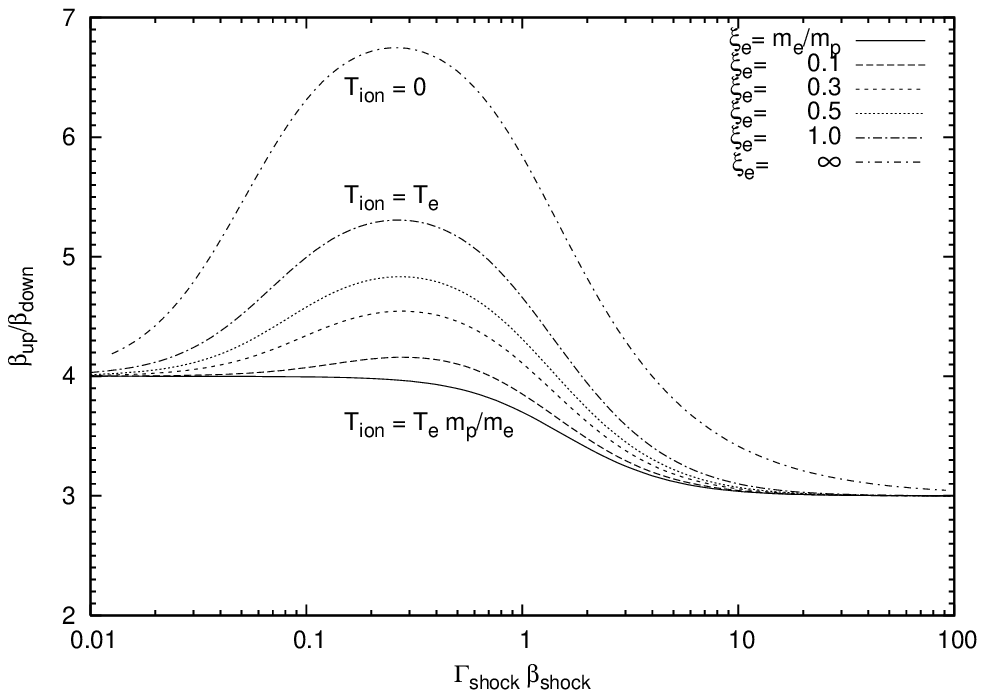}
\caption{Velocity compression factor at the shock when the
downstream protons transfer a fraction $\xi_e$ of their thermal energy to
electrons.}
\label{sez3fig_Eq_stato_p+e}
\end{center}
\end{figure}

We found the appropriate solutions for the jump conditions for several
values of the parameter $\xi_e$ in the range $m_e/m_p \leq \xi_e \leq 1$.  
Fig. \ref{sez3fig_Eq_stato_p+e} shows the velocity ratio $r_\beta$ as a
function of the product $\Gamma_{sh}\beta_{sh}$ for
$\xi_e=m_e/m_p,~0.1,~0.3,~0.5,~1$ and $\infty$. 
The case considered in section \ref{sez3.1} corresponds to $\xi_e = m_e/m_p $,
while $\xi_e = 1$ represents the limit case with protons and electrons
thermalized at the same temperature. 

The most peculiar feature of the curves plotted in
Fig. \ref{sez3fig_Eq_stato_p+e} is the presence of a peak
approximately at the place where the shock becomes trans-relativistic.
The nature of this peak and the phenomenological implications of its
presence are rather interesting. In order to understand the origin of
the peak we consider the unphysical case $\xi_e=\infty$, which
corresponds to completely cold ions in the downstream
plasma. For all curves in Fig. \ref{sez3fig_Eq_stato_p+e} we can
clearly identify three regimes: 1) both electrons and protons
downstream are non-relativistic; 2) electrons are relativistic 
while protons are still non-relativistic; 3) both electrons and 
protons are relativistic. 

The case $\xi_e=m_e/m_p$ (or equivalently $\xi_e\approx 0$) is
the case that is usually studied in the literature. In the limit of non
relativistic strong shocks this case leads to compression factor that
asymptotically approaches 4. The derivative of the compression factor with
respect to $\beta_{sh}$ in this non relativistic regime is zero, as can be
shown by using the Taub conditions Eqs. (\ref{sez2eq_jump_1a})-(\ref{sez2eq_jump_3a})
 (or equivalently, and more easily, the non relativistic version, the Rankine-Hugoniot jump
conditions). On the other hand, if one expands the function $G(x)$ in the
equation of state of electrons to second order in the variable $1/x$
(\textit{i.e.} for mildly relativistic temperature), keeping the protons
non relativistic, the resulting total equation of state reads
\begin{equation} \label{sez3.2eq_peak1}
 \bar\epsilon_T = 1+ \frac{3}{2} \, \bar p_T +\frac{15}{8}
\frac{\xi_e^2}{(1+\xi_e)^2} \, \frac{1}{\bar\rho_e} \, \bar p_T^2 \,.
\end{equation}
It is easy to show that the Rankine-Hugoniot relations give now a
compression factor with a positive derivative with respect to
$\beta_{sh}$, for small values of $\beta_{sh}$.
Moreover the asymptotic value of $r_\beta$ for ultra-relativistic shocks is
always $3$. This implies that at some point in between the compression
factor has a peak where the derivative is zero and the compression factor
is maximum. This is clearly seen in  Fig. \ref{sez3fig_Eq_stato_p+e}: in
the trans-relativistic regime, electrons become relativistic downstream
before protons do (and even if the shock is not fully relativistic) thereby
making the downstream fluid more compressive. To show that this is the
correct interpretation, one can estimates the value of $r_\beta$ at the
peak using the equation of state for non relativistic protons ($\epsilon_p=
\rho_p c^2+3p_p/2$) and fully relativistic electrons ($\epsilon_e= 3 p_e$).
Under this assumption the total normalized energy reads
\begin{equation} \label{sez3.2eq_peak2}
 \bar\epsilon_T = \bar \rho_p + \frac{3}{2} \,
\frac{1+2\xi_e}{1+\xi_e} \, \bar p_T \,.
\end{equation}
Inserting this equation in the Rankine-Hugoniot relations one obtains the
following expression for the compression ratio:
\begin{equation} \label{sez3.2eq_peak3}
 r_\beta = \frac{4+7\,\xi_e}{1+\,\xi_e}\,.
\end{equation}	
Substituting $\xi_e$ with the values listed in Fig.
(\ref{sez3fig_Eq_stato_p+e}), one recovers values of the compression
factor close to that of the peaks within an error $\lesssim 4\%$.

\subsection{Turbulent magnetic field production} 
\label{sez3.3} 

It is interesting to investigate the possibility that part of
the ram pressure of the upstream fluid may be converted into a turbulent
magnetic field in the downstream region. Here we investigate this
scenario and in particular we calculate the compression factor at 
the shock and the spectrum of the accelerated particles.
In order to take into account the dynamical effect of the turbulent
magnetic field it is necessary to generalize the Taub conditions at
the shock, (\ref{sez2eq_jump_2a})-(\ref{sez2eq_jump_3a}), by
introducing the proper components of the electromagnetic stress
tensor $T^{\mu\nu}$. The specific energy density in the form of
turbulent magnetic field is $\epsilon_m= T^{00}$ while the pressure in
the direction identified by the index $i$ is $p_{m,i}= T^{ii}$.

In the plasma reference frame, where no electric field is present, the 
electro-magnetic energy tensor can be written as follows:
\begin{equation}
T^{\mu \nu} = \frac{1}{4\pi} \left(F^{\mu \alpha} F^\nu_\alpha 
         - \eta^{\mu\nu} F^{\alpha \beta} F_{\alpha \beta}\right)
         = -\frac{1}{4\pi} \left( B^\mu B^\nu -\frac{1}{2} B^2
           \eta^{\mu\nu}\right) \,.   \label{sez3.3eq_Tensor} 
\end{equation}
Here $B^{\mu}=(0,\textbf{B})$ and  $B^2=B_x^2 +B_y^2+B_z^2$. To be
included in Eqs. (\ref{sez2eq_jump_2a})-(\ref{sez2eq_jump_3a}) $T^{\mu
\nu}$ has to be expressed in the shock frame, where both the
scaler $B$ and the field component along the propagation direction $B_x$
remain unchanged. Hence the energy density and the pressure along the shock
propagation direction are
\begin{eqnarray} 
 \epsilon_m &=& ( B_x^2 + B_y^2 + B_z^2)/8\pi \,,  \label{sez3.3eq_e_m} \\ 
  p_{m,x}   &=& (-B_x^2 + B_y^2 + B_z^2)/8\pi \,.  \label{sez3.3eq_p_m}
\end{eqnarray}

Two situations may be of interest here: 1) the turbulent field is
created directly behind the shock. In this case the strength of the
field is equally distributed among the three spatial dimensions. 2)
the turbulent field downstream results from the compression of a
turbulent field upstream. In this second case the two components of
the field which are perpendicular to the shock normal are amplified at
the shock while the parallel component is left unaltered. 
In the former case the relation between energy density and pressure is
easily obtained to be
\begin{equation}  \label{sez3.3eq_state_B}
p_m = \epsilon_m / 3 \, ,
\label{eq:case1}
\end{equation} 
where the factor $1/3$ suggests that the magnetic field behaves like a
relativistic gas irrespective of the shock speed. On the other hand,
in the latter case, if the shock is ultra-relativistic, the parallel
component of the turbulent magnetic field is negligible with respect
to the perpendicular components due to the shock compression. If the
parallel component is neglected, the relation between energy and
pressure of the magnetic field is easily obtained to be $p_m =
\epsilon_m$.

In the following we limit ourselves with the case $p_m = \epsilon_m/3$, 
but we discuss the case $p_m = \epsilon_m$ in Sec. \ref{sec:spectrum}.

If $\xi_m$ is the fraction of magnetic energy density with respect to the
proton kinetic energy density, we can write
\begin{equation} \label{sez3eq_Energy_B}
\bar \epsilon_m = \xi_m \,(\bar\epsilon_p - \bar\rho_p )   \, .
\end{equation} 

When the magnetic energy equals the kinetic energy of protons, 
the magnetic pressure is smaller than that of protons in the Newtonian 
limit. On the other hand the two pressures become equal in the 
ultra-relativistic limit. It follows that we expect the compression 
ratio to increase in the Newtonian limit as the magnetic contribution 
increases, while the compression factor levels off when the
relativistic regime is approached. 
The total downstream equation of state when only protons and a
turbulent magnetic field are taken into account is:
\begin{equation} \label{sez3eq_Energy_p+B}
\bar \epsilon_{\rm T} = \bar\epsilon_p + \bar\epsilon_m =\bar\epsilon_p
(1+\xi_m) - \xi_m \bar\rho_p \, .
\end{equation} 

\begin{figure}
\begin{center}
\includegraphics[angle=0,scale=1]{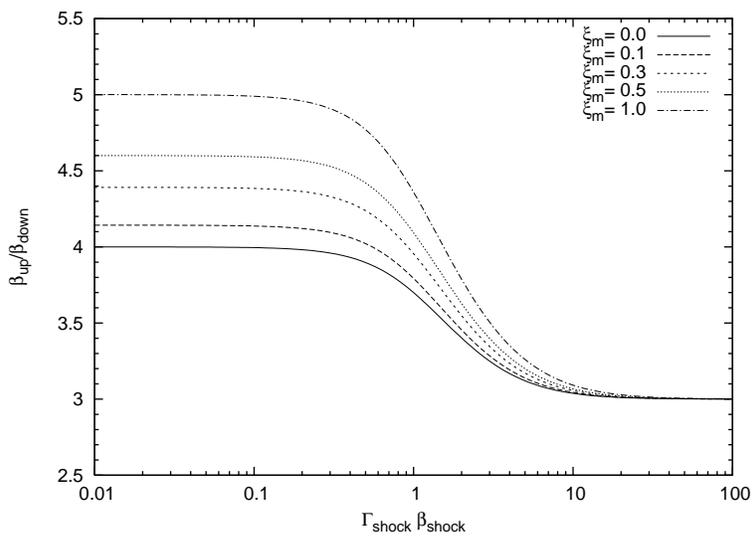}
\caption{Velocity compression factor at the shock when a turbulent
  magnetic field is present inside the electron-proton plasma, and 
$\xi_m$ is the fraction of magnetic field energy
density with respect to the protons kinetic energy.}
\label{sez3fig_Eq_stato_p+B}
\end{center}
\end{figure}

Fig. \ref{sez3fig_Eq_stato_p+B} shows the compression factor $r_\beta$ 
for this situation, for several values of $\xi_m$, in the range $0<\xi_m<1$.
The velocity compression factor ranges from $5.0$ in the Newtonian limit to
$3.0$ in the ultra-relativistic limit, when protons and the magnetic field
are considered in equipartition (\textit{i.e.} $\xi_m=1$).

\subsection{The general case} \label{sez3.4}

As a natural conclusion of our exercise, we consider now the case of
an electron-proton plasma with all the effects introduced above.
The total energy of the system is given by the sum of the Eqs. 
(\ref{sez3eq_Energy_p}), (\ref{sez3eq_Energy_e}) and
(\ref{sez3eq_Energy_B}). The total pressure is the sum of the three
contributions due to protons, electrons and magnetic pressure. We can
write:  
\begin{eqnarray} 
\bar \epsilon_{\rm T}(\bar p_p) 
	&=& \bar\epsilon_p + \bar\epsilon_e + \bar\epsilon_m 
 \nn \\ 
        &=& \bar\epsilon_p (1+\xi_m) + \bar\epsilon_e -
 \bar \rho_p \xi_m   \nn   \\
	&=& (1+\xi_m) \bar \rho_p \, G{\left( \frac{\bar \rho_p}{\bar p_p}
\right)} + \bar \rho_e \, G{\left( \frac{\bar \rho_e}{\xi_e \bar p_p}
\right)}  \nn \\
	& & -\bar p_p (1+ \xi_m + \xi_e) - \bar \rho_p \xi_m \,
,  \label{sez3eq_Energy_p+e+B+loss}  \\
\bar p_{\rm T}(\bar p_p) &=& \bar p_p + \bar p_e +\bar p_m   \nn  \\ 
	&=& \bar p_p \left(1+\xi_e-\frac{\xi_m}{3} \right) +
\frac{\xi_m}{3}
\bar \rho_p \, \left[G{\left( \frac{\bar \rho_p}{\bar p_p} \right)} -1
\right] \, ,
\label{sez3eq_Pressure_p+e+B+loss}
\end{eqnarray} 
where we also introduced the proton and electron normalized
densities $\bar \rho_p= m_p/(m_p+m_e)$ and $\bar \rho_e= m_e/(m_p+m_e)$.

\begin{figure}
\begin{center}
\includegraphics[angle=0,scale=1]{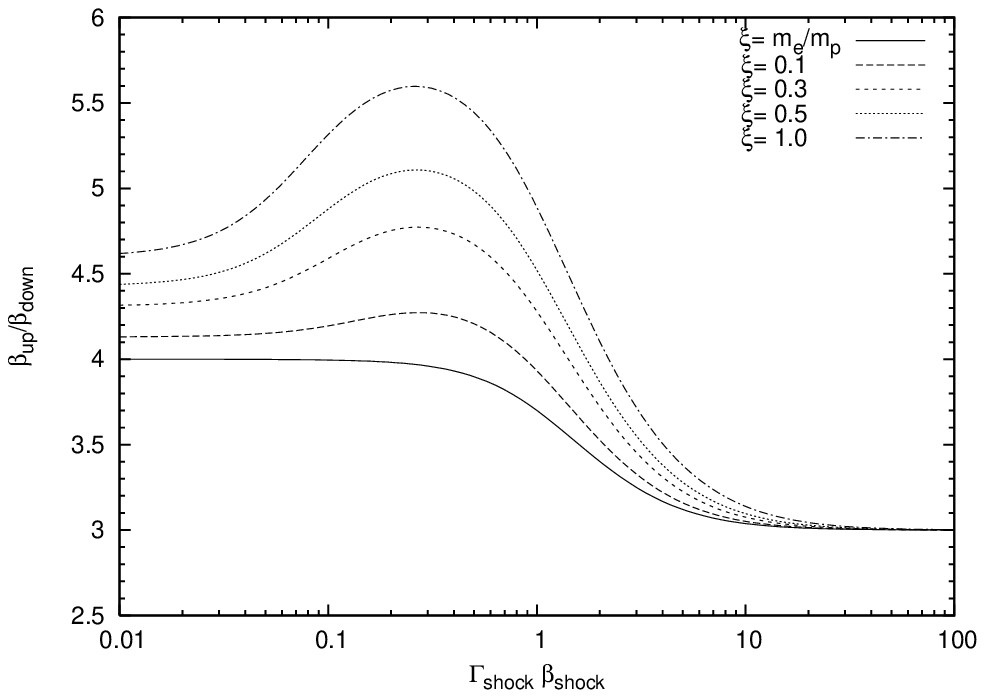}
\caption{Velocity compression factor when the equation of state of the
  downstream plasma takes into account the exchange of energy between
  electrons and protons, and the generation of a turbulent magnetic field.
  Here we assumed $\xi_e = \xi_m \equiv \xi$.}
\label{sez3fig_Eq_stato_p+e+B+loss}
\end{center}
\end{figure}

In order to avoid the effects of proliferation of free parameters, in
the following we limit ourselves with a sort of equipartition
situation, in which $\xi_e=\xi_m \equiv \xi$. 
Fig. \ref{sez3fig_Eq_stato_p+e+B+loss} shows the resulting compression
factor $r_\beta$ for $m_e/m_p \leq \xi \leq 1$. 

It is important to stress that for shocks in the newtonian and
trans-relativistic regime, the magnetic field and the thermal
electrons both result in making the plasma more compressible (the
compression factor is as high as $5.6$ at the peak 
$\Gamma_{sh}\beta_{sh}\simeq 0.3$, when equipartition $\xi=1$ is
assumed). On the other hand for highly relativistic shocks all the three
components behave in the same way resulting in a compression factor equal
to 3. The electron contribution turns out to be especially important in the
intermediate velocity range ($0.1\lesssim \Gamma_s \beta_s \lesssim 1$).

\section{The particles' spectrum}
\label{sec:spectrum}

The spectrum and angular distribution of the particles accelerated at
shocks with arbitrary speed and arbitrary scattering properties of the
fluid can be calculated following the theory of particle
acceleration put forward by \cite{vie03}. This approach requires the
calculation of the conditional probabilities of a particle returning
from upstream or downstream at some direction given the entrance
direction. These conditional probabilities were described by
\cite{bla05} in terms of two non-linear integral equations, that were
solved iteratively. 

The theoretical approach of \cite{vie03} and \cite{bla05} applies
equally well to cases of Small Pitch Angle Scattering (SPAS) and Large
Angle Scattering (LAS), and to the case of a large scale coherent
field upstream \cite{mor06}. Here we limit ourselves with considering
only two situations, namely that of SPAS, both upstream and
downstream, and that of a large scale field upstream, with orientation
perpendicular to the shock normal (perpendicular shock). In
\cite{mor06} it was shown that the spectral shape does not change
dramatically with the inclination, with the exception of the cases
where the shock is quasi-parallel, but these cases lead to
insignificant acceleration and are therefore physically irrelevant.

\begin{figure}
\begin{center}
\includegraphics[angle=0,scale=1]{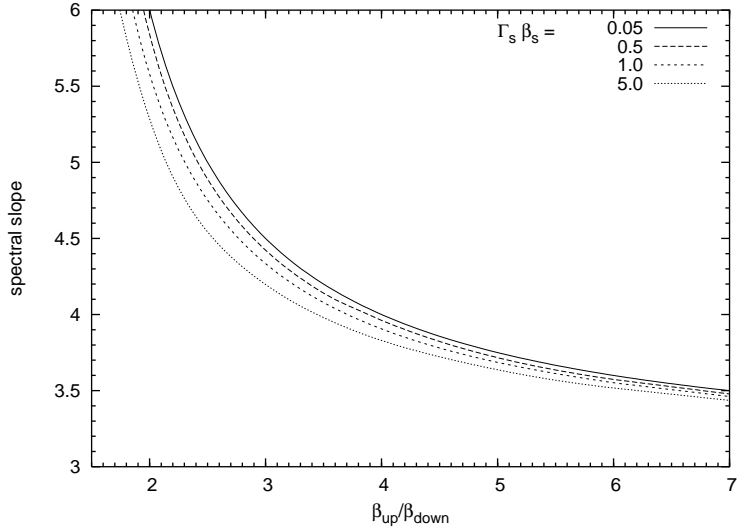}
\caption{Slope of the spectrum of accelerated particles as a function
  of the velocity compression factor for different values of the shock
  speed. The scattering in the SPAS regime both upstream and downstream.}
\label{sez4fig_slope_vs_r}
\end{center}
\end{figure}

\begin{figure}
\begin{center}
\plottwo{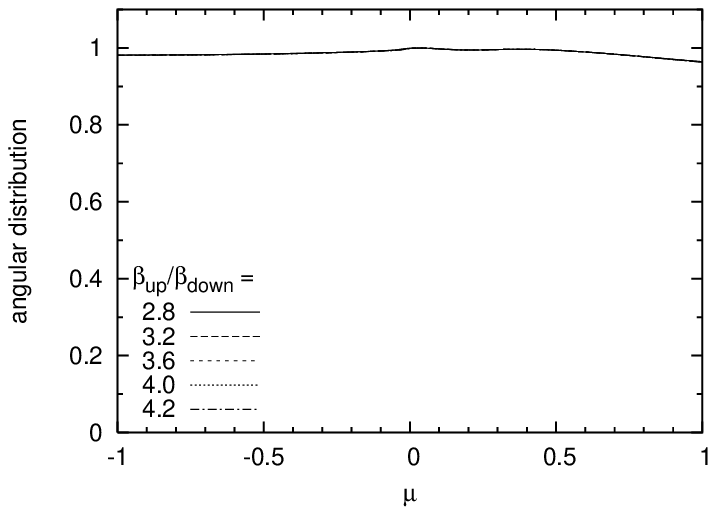}{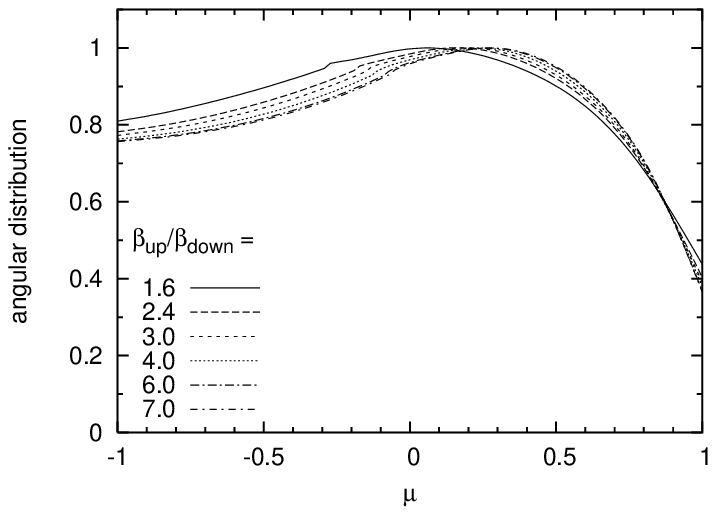}
\plottwo{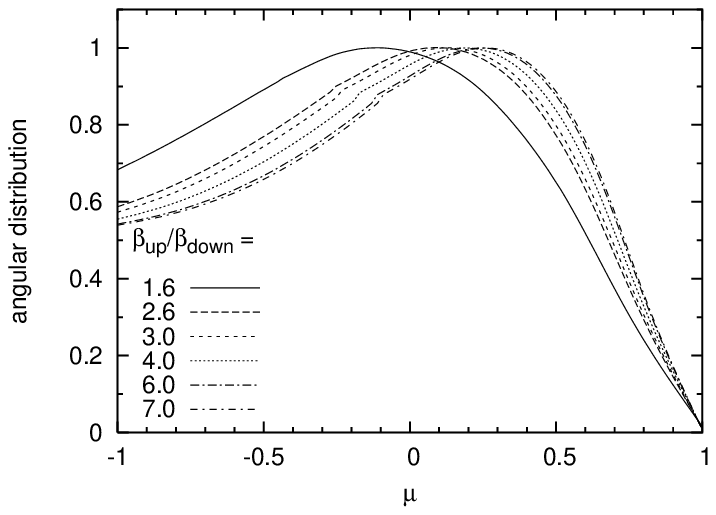}{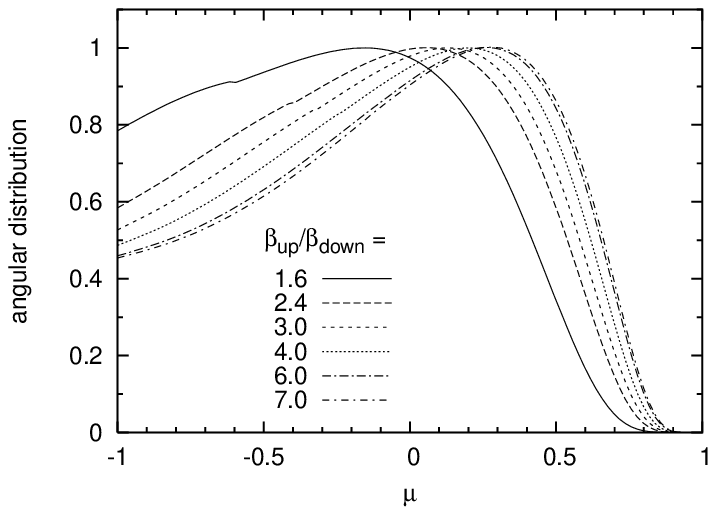}
\caption{Angular distribution of accelerated particles at the shock front as
measured in the downstream fluid frame. $\mu$ is the cosine of the angle
between the particle direction and the shock normal. The different
lines correspond to different values of the velocity compression
factor, as indicated in the labels, while the shock velocity has a 
fixed value for each plot: $\Gamma_{sh} \beta_{sh}=0.1$ (top-left), $0.5$
(top-right), $1.0$ (bottom-left) and $5.0$ (bottom-right).}
\label{sez4fig_ang_dis}
\end{center}
\end{figure}

Before showing our results for the different equations of state
discussed above, it is useful to show the spectral slope and the
distribution function of the accelerated particles for different
values of the velocity compression factor. In
Fig. \ref{sez4fig_slope_vs_r} we show the spectral slope as a function
of the compression factor for several values of
$\Gamma_{sh}\beta_{sh}$, ranging from $0.05$ to $5.0$. The solid line 
corresponds to the case of a strong newtonian shock, $s(r_\beta) =
(r_\beta+2)/(r_\beta-1)$. In Fig. \ref{sez4fig_ang_dis} we also show
the distribution function of the accelerated particles as a function
of the direction $\mu$ measured in the downstream frame. 

As discussed in \S\S \ref{sez3.1}-\ref{sez3.4}, the basic effect of
changing the equation of state is to change the compression factor at 
the shock and thereby the shape of the spectrum of the accelerated
particles. In the following we discuss separately the effect of the
interaction between electrons and protons in the downstream plasma, 
and the effect of the generation of magnetic field. Finally we shall
use Eqs. (\ref{sez3eq_Energy_p+e+B+loss}) and
(\ref{sez3eq_Pressure_p+e+B+loss}) in order to quantify the combined
effect of the two phenomena.

\begin{figure}
\begin{center}
\plottwo{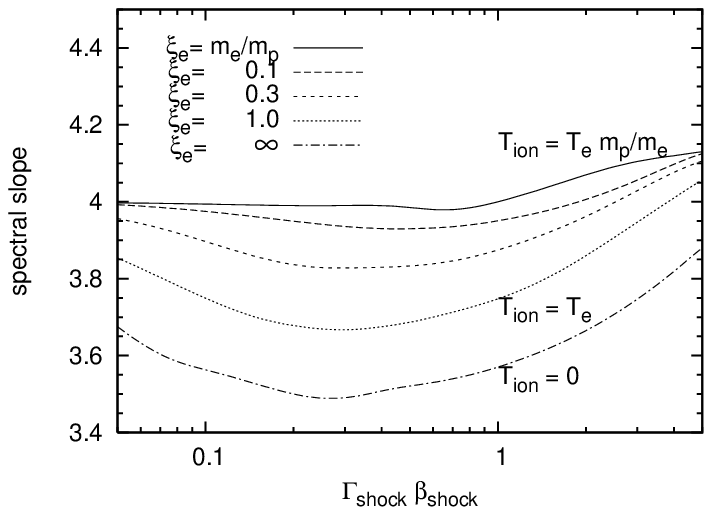}{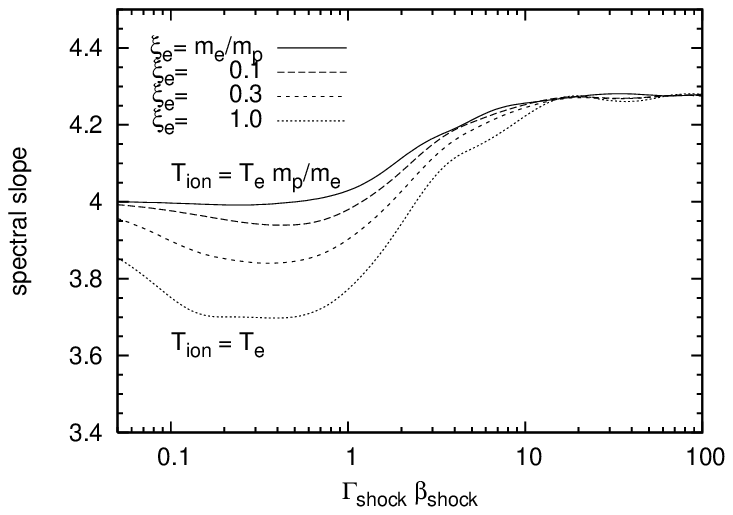}
\caption{Slope of the spectrum of accelerated particles as a function
  of $\beta_{sh}\Gamma_{sh}$ when the electrons have a temperature
  $T_e= \xi_e T_p$. The left panel refers to the case of SPAS in both
  the upstream and downstream plasmas. The right panel refers to the
  case in which a regular perpendicular field is present upstream.}
\label{sez4fig_slope_p+e}
\end{center}
\end{figure}

Fig. \ref{sez4fig_slope_p+e} shows the spectral slope as a function of
the shock speed for different values of the parameter $\xi_e$ which
characterizes the degree of coupling between thermal electrons and
thermal protons, as introduced in \S\S \ref{sez3.1}. The left panel
refers to the case of SPAS both upstream and downstream, while the
right panel refers to the case of a regular perpendicular field
upstream (and SPAS downstream). 

Some comments are in order: for very non-relativistic shocks (not
shown in the plot) the spectrum has the standard slope $s=4$. In the
ultra-relativistic limit universality is also reached, being the
results independent on the value of $\xi_e$. Most differences in the
slope of the spectrum of accelerated particles is present for
trans-relativistic shocks: a minimum in the slope appears for these
shocks, deeper for larger values of $\xi_e$. This flattening
of the spectrum is due to the increasing compression ratio at
trans-relativistic shock speed, as explained in the last paragraph of \S
\ref{sez3.2}.  A very similar behaviour was found also in \cite{kir99}
(compare their Fig 3 with left panel of our Fig. \ref{sez4fig_slope_p+e})
where they use a different technique, the eigenvalue expansion firstly
introduced by \cite{kir87}, and an equation of state for a gas consisting
of both hydrogen and helium.

\begin{figure}
\begin{center}
\plottwo{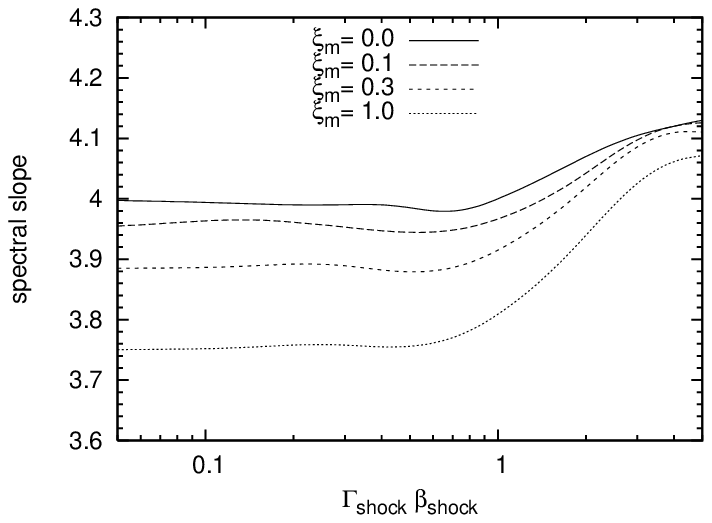}{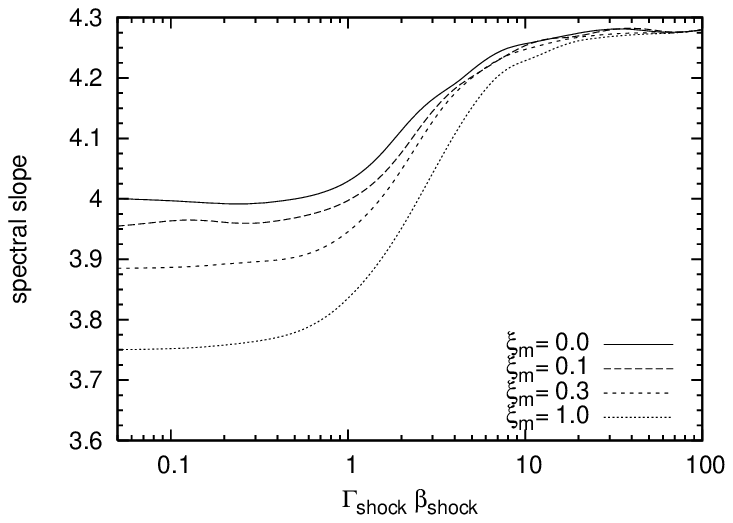}
\caption{Slope of the spectrum of accelerated particles as a function
  of $\beta_{sh}\Gamma_{sh}$ when a turbulent magnetic field is
  present with $\epsilon_m=\xi_m (\epsilon_p-\rho_p c^2)$. 
  The left panel refers to the case of SPAS in both
  the upstream and downstream plasmas. The right panel refers to the
  case in which a regular perpendicular field is present upstream.}
\label{sez4fig_slope_p+B}
\end{center}
\end{figure}

Based on the discussion in \S\ref{sez3.3}, the effect of a
turbulent magnetic field is expected to depend upon the shock speed. As
illustrated in Fig. \ref{sez4fig_slope_p+B} (left and right panel as
in Fig. \ref{sez4fig_slope_p+e}) the spectrum of accelerated particles 
is harder than in the absence of magnetic fields for newtonian and
trans-relativistic shocks (the minimum slope is $s=3.75$ in the 
equipartition regime, $\xi_m\sim 1$). In the ultra-relativistic regime 
all the curves approach the same value: the configuration of magnetic 
field adopted here does not produce any change in the particle spectra 
with respect to the case where no magnetic field is present. On the 
other hand this is easy to guess simply looking at Eq. 
(\ref{sez3.3eq_state_B}).

The spectrum of the accelerated particles is however rather
sensitive to the structure of the turbulent field in the downstream
fluid. As we discussed in Sec. \ref{sez3.3}, the equation of state of the
turbulent field depends on whether the field is generated downstream
and is therefore isotropic in the local frame, or it is rather
compressed in its perpendicular components. In this latter case the
equation of state of the field is not the same as that of a
relativistic fluid and this affects the compression factor at the
shock. As an instance we consider here the case
$\Gamma_{sh}\beta_{sh}=5$ and we introduce a magnetization parameter 
$\alpha=(\delta B^2/8\pi)/\rho c^2$ in the upstream region, where $\delta
B$ is the amplitude of the average magnetic turbulence. $\alpha=0$
corresponds to the unmagnetized case, which leads to a compression factor
$3.12$ and a spectral slope $s=4.12$, as already found earlier. On the
other hand, for $\alpha=10^{-2}$ ($\alpha=3\times 10^{-2}$) the compression
factor becomes $2.71$ ($2.16$) and the spectral slope is $s=4.4$
($s=4.95$). The corresponding value of the parameter $\xi_m$
downstream, as resulting from the compression of the perpendicular
components of the magnetic field is $\xi_m=0.11$ ($\xi_m=0.28$).
This softening of the spectrum may have very important
phenomenological consequences for those classes of sources where
particle acceleration occurs at ultra-relativistic shocks.

\begin{figure}
\begin{center}
\plottwo{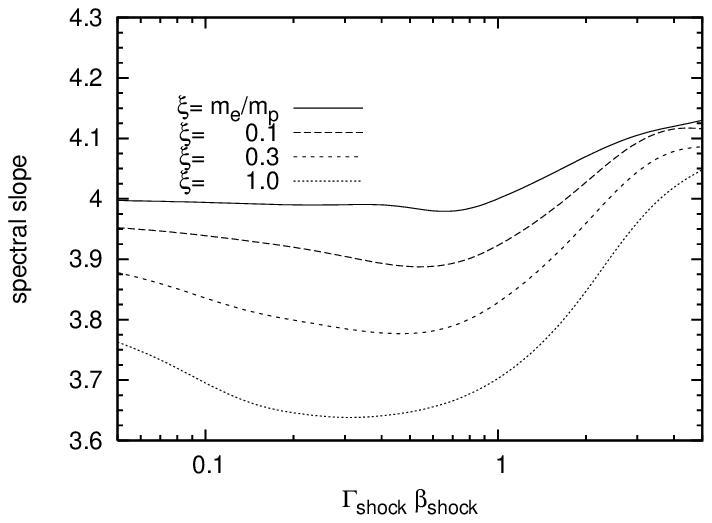}{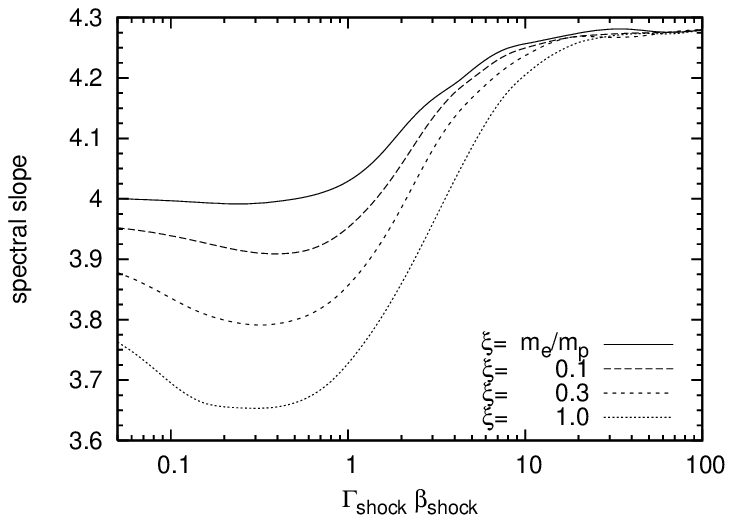}
\caption{Slope of the spectrum of accelerated particles as a function
  of $\beta_{sh}\Gamma_{sh}$ with the full equation of state for the
  downstream plasma.
  The left panel refers to the case of SPAS in both
  the upstream and downstream plasmas. The right panel refers to the
  case in which a regular perpendicular field is present upstream.}
\label{sez4fig_slope_p+e+B}
\end{center}
\end{figure}

Finally we consider the case in which both the effects of turbulent
magnetic field downstream and energy exchange between the thermal
components of electrons and protons are taken into account. More
specifically we concentrate on the so called {\it equipartition case}, in
which $\xi_e=\xi_m=\xi$ and we illustrate our results for different values
of $\xi$. Here we restrict our attention to the case in which the
turbulent field is generated downstream and does not result from the
compression of an upstream field. As usual, the left panel in 
Fig. \ref{sez4fig_slope_p+e+B} refers to SPAS both upstream and 
downstream and the right panel to a large scale field upstream (and
SPAS downstream).  

For $\xi=m_e/m_p$ the standard result is recovered in the SPAS
case (left panel). The case $\xi=0$  when a large scale magnetic field 
is present upstream was discussed in \cite{mor06} and the results are
very close to the case $\xi=m_e/m_p$ used as the lower limit in Fig. 
\ref{sez4fig_slope_p+e+B}.

In the other extreme case $\xi=1$ the spectrum of accelerated
particles is harder than $4$ in the non-relativistic limit and
approaches the {\it universal} spectrum in the ultra-relativistic
case. In both cases of SPAS and regular field upstream, a minimum in
the slope of the spectrum is reached at $\beta_{sh}\Gamma_{sh}\approx
0.4$ corresponding to $s\approx 3.65$. This is a result of the
interaction between electrons and protons in the trans-relativistic
regime.

In Figs. \ref{sez4fig_slope_p+e}, \ref{sez4fig_slope_p+B}
and \ref{sez4fig_slope_p+e+B} the range of shock velocities adopted
for the cases of upstream ordered magnetic field and of SPAS are quite
different. This is not due to any deep reason, but simply to the fact
that the computation method introduced in \cite{bla05} becomes very
challenging when the Lorentz factor becomes too large. This is related
to the fact that in order to describe the regime of small pitch angle
scattering we need to adopt a finite aperture $\sigma$ of the
scattering function which however must stay smaller than
$1/4\Gamma_{sh}^2$ in order to operate in the SPAS regime. As
discussed in \cite{mor06} this problem is 
absent in the case of ordered magnetic field.

\section{Conclusions}
\label{sec:conclusions}

We used the theoretical framework introduced by \cite{vie03} and 
\cite{bla05} to determine the spectrum of particles accelerated at
shocks with arbitrary speed and scattering properties, with different
equations of state of the gas downstream. In particular we investigated  
two situations: 1) the downstream gas is made of thermal electrons and
protons that may exchange energy with each other, thereby changing the
equation of state; 2) the downstream gas includes a turbulent magnetic
field. We also considered the case in which both effects are at work
at the same time. 

In the downstream frame the scattering has always been assumed to be
in the SPAS regime, while in the upstream fluid we considered two
scenarios, namely SPAS and regular field with no turbulent
scattering. We limited our attention to the case of strong shocks,
namely the case in which the inflowing plasma has zero pressure. 

As a result of our calculations, we found several instances of
violation of the so-called universality of the spectrum in both the
newtonian and the trans-relativistic regimes. 

When the downstream plasma is made of electrons and protons and their
temperatures are different, for newtonian and ultra-relativistic shocks
the shape of the spectrum is not appreciably changed. However, for
trans-relativistic shocks the compression factor increases appreciably
when the electron temperature is $\gg \frac{m_e}{m_p}T_p$, causing a
flattening in the spectrum of the accelerated particles. In particular
this is true for $T_e=T_p$, a situation which might be achieved due to
some efficient collisionless plasma process, able to equilibrate the
electron and proton components more efficiently than the simple
isotropization of the velocity vectors. 

The effect of a turbulent magnetic field on the compression
factor and on the spectral slope is more complex. We identified two
situations of interest that arise when the magnetic field is
introduced in the conservation equations at the shock surface: 1)
the turbulent field is created downstream and is isotropic; 2) the
magnetic field downstream is the result of the compression of the
turbulent field upstream (only the perpendicular components are
compressed). In the former case it can be demonstrated that the
equation of state of the magnetic field is identical to that of an
ideal relativistic gas, Eq. (\ref{eq:case1}), irrespective of the shock
speed. When the shock is non-relativistic the compression factor at the
shock is increased and the spectra become harder than in the absence of
field. For relativistic shocks the usual asymptotic spectrum is reached
both in the case of upstream regular deflection or small pitch
angle scattering.

The second case is more interesting: when the perpendicular
components of the turbulent upstream field are amplified by crossing
the shock surface, the resulting magnetic field downstream is strongly
anisotropic and the equation of state of the magnetic field is
$p_m=\epsilon_m$, not resembling that of a relativistic gas. In this
case, for relativistic shocks the spectra of accelerated particles are
softer than in the first scenario. For instance for
$\Gamma_{sh}\beta_{sh}=5$ and $\sigma=10^{-2}$ ($\sigma=3\times
10^{-2}$) the compression factor becomes $2.71$ ($2.16$) and the
spectral slope is $s=4.4$ ($s=4.95$). The corresponding value of the
parameter $\xi_m$ downstream, as resulting from the compression of the
perpendicular components of the magnetic field is $\xi_m=0.11$
($\xi_m=0.28$). 
The important effect consisting of a spectral steepening was found by
\cite{lem06} and was attributed to the fact that the compression in
the downstream gas makes the magnetic field quasi-perpendicular,
thereby reducing significantly the probability of return from the
downstream frame. This effect is limited, in the analysis of
\cite{lem06} to the particles with gyration radius smaller than the
coherence scale of the turbulent field. The steepening of the spectrum
as found in our calculations, is due to a change in the equation of
state of the downstream plasma (electrons, protons and magnetic field)
and concerns all of the spectrum of the accelerated particles.
The relevance of this finding for the phenomenology of several
astrophysical sources of accelerated particles, in particular those
where a relativistic shock is expected or observed, is evident.

\acknowledgments{The authors are grateful to an anonymous referee for
  the very interesting comments that helped us improving the
  manuscript considerably. This research was partially funded through
  grant Prin-2004.}

\clearpage


\begin{thebibliography}{}

\bibitem[Achterberg et al. 2001]{ach01}
Achterberg, A., Gallant, Y.A., Kirk, J.G., \& Guthmann, A.W. 2001, \mnras,
328, 393

\bibitem[Amato \& Blasi 2005]{amato1}
Amato, E., and Blasi, P., 2005, MNRAS Lett., 364, 76

\bibitem[Amato \& Blasi 2006]{amato2}
Amato, E., and Blasi, P., 2006, MNRAS {\it in press}, Preprint
astro-ph/0606592 

\bibitem[Bednarz \& Ostrowski 1998]{bed98}
Bednarz, J., \& Ostrowski, M. 1998, \prl, 80, 3911

\bibitem[Begelman \& Chiueh 1988]{beg88} 
Begelman, M.C., \& Chiueh, T. 1988, \apj, 332, 827-890

\bibitem[Bell 1978a]{bell78a}
Bell, A.R., 1978a, MNRAS, 182, 147

\bibitem[Bell 1978b]{bell78b}
Bell, A.R., 1978b, MNRAS, 182, 443

\bibitem[Bell 1987]{bell87}
Bell, A.R., 1987, MNRAS, 225, 615

\bibitem[Blandford \& Ostriker 1978]{bo78}
Blandford, R.D. and Ostriker, J.P., 1978, ApJL, 221, 29

\bibitem[Blasi 2002]{blasi1}
Blasi, P., 2002, Astropart. Phys. 16, 429

\bibitem[Blasi 2004]{blasi2}
Blasi, P., 2004, Astropart. Phys. 21, 45

\bibitem[Blasi et al. 2005]{vannoni}
Blasi, P., Gabici, S., and Vannoni, G., 2005, MNRAS, 361, 907 

\bibitem[Blasi \& Vietri 2005]{bla05} 
Blasi, P., \& Vietri, M., 2005, \apj, 626, 877

\bibitem[Drury \& V\"{o}lk 1980]{dr_v80}
Drury, L.O'C and V\"{o}lk, H.J., 1980, Proc. IAU Symp. 94, 363

\bibitem[Drury \& V\"{o}lk 1981]{dr_v81}
Drury, L.O'C and V\"{o}lk, H.J., 1981, ApJ, 248, 344

\bibitem[Ellison et al. 1995]{ebj95}
Ellison, D.C., Baring, M.G., and Jones, F.C., 1995, ApJ, 453, 873

\bibitem[Ellison et al. 1996]{ebj96}
Ellison, D.C., Baring, M.G., and Jones, F.C., 1996, ApJ, 473, 1029

\bibitem[Ellison \& Double 2002]{double}
Ellison, D.C., and Double, G.P., 2002, Astropart. Phys., 18, 213

\bibitem[Ellison et al. 1990]{elli90}
Ellison, D.C., M\"{o}bius, E., and Paschmann, G., 1990, ApJ, 352, 376

\bibitem[Gallant 2002]{gal02} 
Gallant, Y.A. 2002, Relativistic flow in Astrophysics, LNP 589, 24

\bibitem[Gallant \& Achterberg 1999]{gal99}
Gallant, Y.A., \& Achterberg, A. 1999, \mnras, 305, L6
 
\bibitem[Gallant et al. 1992]{gal92}
Gallant, Y.A., Hoshino, M., Langdon, A.B., Arons, J., \& Max, C.E. 1992, 
\apj, 391, 73

\bibitem[Heavens \& Drury 1988]{heavens}
Heavens, A.F., and Drury, L.O'C., 1988, \mnras, 235, 997

\bibitem[Hoshino et al. 1992]{hos92} 
Hoshino, M.H., Arons, J., Gallant, Y.A., \& Langdon, A.B. 1992, \apj, 390,
454

\bibitem[Jones \& Ellison 1991]{je91}
Jones, F.C. and Ellison, D.C., 1991, Space Sci. Rev. 58, 259

\bibitem[Kang \& Jones 1997]{kj97}
Kang, H., and Jones, T.W., 1997, ApJ, 476, 875

\bibitem[Kang \& Jones 2005]{kj05}
Kang, H., and Jones, T.W., 2005, ApJ, 620, 44

\bibitem[Kang et al. 2002]{jones02}
Kang, H., Jones, T.W., and Gieseler, U.D.J., 2002, ApJ 579, 337

\bibitem[Kirk \& Duffy 1999]{kir99}
Kirk, J.G., and Duffy, P. 1999, J. Phys. G., 25, R163

\bibitem[Kirk et al. 2000]{kir00}
Kirk, J.G., Guthmann, A.W., Gallant, Y.A., \& Achterberg, A., 
2000, ApJ 542, 235

\bibitem[Kirk \& Schneider 1987]{kir87}
Kirk, J.G., and Schneider, P. 1987, \apj, 315, 425

\bibitem[Krymskii 1977]{krymskii}
Krymskii, G.F., 1977, Soviet Physics - Doklady 327, 328

\bibitem[Lemoine \& Pelletier 2003]{lem03}
Lemoine, M., \& Pelletier, G. 2003, \apj, 589, L73

\bibitem[Lemoine \& Revenu 2006]{lem06}
Lemoine, M., \& Revenu, B. 2006, \mnras, 366, 635

\bibitem[Malkov 1997]{malkov1}
Malkov, M.A., 1997, ApJ, 485, 638

\bibitem[Malkov et al. 2000]{malkov2}
Malkov, M.A., Diamond P.H., and V\"{o}lk, H.J., 2000, ApJL, 533, 171 

\bibitem[Malkov \& Drury 2001]{drurymalkov}
Malkov, M.A. and Drury, L.O'C., 2001, Rep. Prog. Phys. 64, 429

\bibitem[Morlino et al. 2006]{mor06} 
Morlino, G., Blasi, P., \& Vietri, M. 2006, \apj({\it in press}),
Preprint astro-ph/0701173

\bibitem[Niemiec \& Ostrowski 2004]{nie04}
Niemiec, J., \& Ostrowski, M. 2004, \apj, 610, 851

\bibitem[Peacock 1981]{peacock}
Peacock, J.A., 1981, \mnras, 196, 135

\bibitem[Synge 1957]{syn57} 
Synge, J.L. 1957, The relativistic gas (Amsterdam: North-Holland)

\bibitem[Vietri 2003]{vie03} 
Vietri, M. 2003, \apj, 591, 954

\end{thebibliography}
\end{document}